\documentstyle[11pt,newpasp,twoside,epsf]{article}
 \def\ltsim{\lower.5ex\hbox{$\; \buildrel < \over \sim \;$}}

\def\edcomment#1{\iffalse\marginpar{\raggedright\sl#1\/}\else\relax\fi}
\reversemarginpar

\begin{document}
\title{Some Comments related to AGN Radio Loudness}
\author{Ari Laor}
\affil{Physics Department, Technion, Haifa 32000, Israel}

\begin{abstract}
The bimodality of the AGN radio loudness distribution, and the relation
of radio loudness and black hole mass were recently disputed. A closer look
at the existing data suggests possible resolutions of these disputes,
as further described below.\footnote {This comment is posted only on astro-ph,
it will not appear in the conference proceedings.}


\end{abstract}

\section{Introduction}
This is an expanded version of comments made during the
recent meeting on ``AGN Physics with the SDSS" (Princeton, July 2003),
concerning the AGN radio
loudness bimodality and the relation of radio loudness and black hole mass.
These comments are not based on my earlier or ongoing work, and rely only
on the existing literature. They are presented here with the hope of
stimulating further work on these subjects.

\section{The Radio Loudness Bimodality of AGN}

The existence of a bimodality in the radio to optical flux ratio ($R$) distribution
in AGN was long thought to be a well established ``textbook" result
(e.g. Peterson 1997; Krolik 1999; Kembhavi \& Narlikar 1999). However,
the recent FIRST survey, a large area, deep, high resolution radio survey
revealed no $R$ bimodality (White et al. 2000), but rather a smooth distribution
with a peak, instead of a dip, at intermediate $R$ values. This surprising
result prompted various attempts to explain the sharp disagreement
with earlier studies of radio and optical emission in AGN
(Ivezi{\' c}~et al.\ 2002; Cirasuolo et al. 2003).

The FIRST survey differs from most earlier surveys
in its significantly higher sensitivity and angular resolution
(Becker, White, \& Helfand 1995). Blundell
(2003) has recently shown a dramatic (but possibly anecdotal) example
of the insensitivity of the FIRST survey images to extended radio emission,
when compared to the lower resolution images of the NVSS survey.
Below I point out existing evidence that the lack of sensitivity to
extended emission may be related to
the lack of $R$ bimodality in the FIRST survey.

Xu, Livio \& Baum (1999) compiled 5~GHz and [O~III] luminosities for
a large and heterogeneous sample of 409 AGN, covering 7 orders of
magnitude in $L_{\rm [O III]}$, and 9 orders of magnitude in $L_{\rm 5 GHz}$.
In low luminosity AGN the optical continuum can be obscured, or heavily
contaminated by the host starlight, and thus $L_{\rm [O III]}$ serves as
a useful proxy for the AGN optical continuum luminosity. Figure 1 in
Xu et al. reveals a clearly bimodal distribution of AGN in the $L_{\rm 5 GHz}$ vs.
$L_{\rm [O III]}$ plane. Radio loud (RL) AGN are on average
about $10^4$ times more radio luminous than radio quiet (RQ) AGN at a given
$L_{\rm [O III]}$. However, a different distribution is seen in their
Figure 3 where they show a plot of the {\em core} 5~GHz
luminosity vs. $L_{\rm [O III]}$ for the same sample. This plot shows
{\em no bimodality}, but rather a smooth distribution of $L_{\rm 5 GHz}$
values at a given $L_{\rm [O III]}$. Thus, the bimodality of the $R$
values distribution in AGN may be true only for the {\em total}, rather
than just {\em core} radio emission. The lack of bimodality in the FIRST
survey may then simply reflect its lack of sensitivity to extended radio
flux, and it may not necessarily contradict the notion of $R$ bimodality in
AGN.

The core radio emission is susceptible to beaming effects, and may
not provide a reliable estimate of the true jet power, unlike the extended
radio emission, which is most likely isotropic and should provide a better
estimate of the jet power. Thus, the bimodality revealed when the extended radio
emission is included is likely to reflect a true bimodality in the fraction
of AGN power emitted in the radio.

\section{The Nature of low $M_{\rm BH}$ Radio Loud AGN}

Various studies of AGN suggested there is a relation between the black hole mass
$M_{\rm BH}$ and radio power (e.g. Franceschini, Vercellone, \& Fabian 1998), or
$M_{\rm BH}$ and radio loudness (e.g. Laor 2000),
such that AGN with $M_{\rm BH}>10^9M_{\odot}$ are all RL, and AGN with
$M_{\rm BH}<10^8M_{\odot}$ are all RQ. However, other studies (Ho 2002;
Oshlack, Webster, \& Whiting 2002; Woo \& Urry 2002) found contradicting
results. In particular, they found a
significant number of low $M_{\rm BH}$ radio loud AGN. As shown below, a closer
inspection of the available data reveals that the published low $M_{\rm BH}$
RL AGN can be classified under
three categories: 1. Wrong $M_{\rm BH}$ determinations.
2. Objects where beaming is likely (and may therfore be intrinsically radio quiet).
3. Very low luminosity AGN. In addition, there
is evidence in the literature
that the $R$ parameter which separates RL and RQ AGN does
not have a fixed value, but rather increases with decreasing luminosity. This
implies that the apparently RL, low $M_{\rm BH}$, low luminosity AGN
may belong to the RQ population. Overall,
the results presented below indicate that there is no robust
case of a luminous RL AGN at $M_{\rm BH}<10^8M_{\odot}$,
and there may be no low $M_{\rm BH}$ ``true" RL AGN at lower luminosities as well.

\subsection{The Woo \& Urry sample}

Woo \& Urry (2002) compiled data from various earlier
studies and find a number of RLQ at $M_{\rm BH}<10^8M_{\odot}$. In their
Table 3 they compiled $M_{\rm BH}$ estimates based on the size of the Broad Line
Region ($R_{\rm BLR}$) and the H$\beta$ FWHM, where the $R_{\rm BLR}$
determination is based on reverberation mappings. They find two RL AGN with
$M_{\rm BH}<10^8M_{\odot}$, 3C120, and PG~1226+230 (3C273).
In 3C120 the $M_{\rm BH}=2.6\times 10^7M_{\odot}$
determination appears quite robust (rather well determined $R_{\rm BLR}$
from reverberation mapping + high quality H$\beta$ optical spectra).
However, this object has a compact and flat spectrum radio source,
it shows superluminal motion, rapid variability and a high polarization,
all of which indicate the radio emission is strongly beamed. In 3C273
the $M_{\rm BH}=1.7\times 10^7M_{\odot}$ value cited from Kaspi
et al. (2000) is erroneous. The actual value in Kaspi et al. is
$(2.4-5.5)\times 10^8M_{\odot}$.

In Table 4 of Woo \& Urry they list $M_{\rm BH}$ determinations
where $R_{\rm BLR}$ is estimated based on the continuum luminosity.
Practically all of the RLQ at $M_{\rm BH}<10^8M_{\odot}$ in this table come
from two studies, Gu, Cao, \& Jiang (2001), and
Oshlack, Webster, \& Whiting (2002) which is discussed in the
following section. Gu et al. find 7
RLQ at $M_{\rm BH}<10^8M_{\odot}$, based on spectra published
and measured by Stickel \& Kuhr (1993a,b),
Stickel, Kuhr, \& Fried (1993), and Lawrence et al. (1996).
The spectra from these studies are enclosed in the Appendix.
In all cases the H$\beta$ line is very narrow, and its FWHM is similar
to the FWHM of the forbidden lines. In addition,
the [O III]$\lambda5007$/H$\beta$ flux ratio is $\sim 10$
(see tables in the above Stickel et al. and Lawrence et al.
papers). These two properties of the measured H$\beta$ component
indicate that it originates purely
from the NLR, and not from the BLR, as assumed by Gu et al.
Since the NLR velocity
dispersion is typically $\sim 10$ smaller than in the BLR,
the BH mass estimate in these objects is a factor of
$\sim 100$ too low.

Some of these 7 objects are not pure type 2 AGN, as they show
some evidence for a BLR component underlying the NLR component.
In 1045$-$188 a broad base of H$\beta$ appears to be partly hidden by
atmospheric absorption, in 1945+726
a very broad component is clearly seen in H$\alpha$, and
in 2218+395 a weak broad base is seen in H$\beta$. However, only
the NLR H$\beta$ component was measured in these objects in the
papers mentioned above.

\subsection{The Oshlack et al. sample}

The Oshlack et al. sample is drawn from the Parkes Half-Jansky
Flat-Spectrum sample (PHFS). As Oshlack et al. comment, the flat radio
spectrum criterion should favor selection of beamed core dominated
objects, and evidence that this indeed happens is provided by
the significant contribution of a steep power law component,
apparently synchrotron emission, to the optical IR emission in
a large fraction of their objects. Apart from the beaming correction
which may significantly reduce the implied intrinsic radio
power (Jarvis \& McLure 2002), there may also be some inaccurate
deductions from the optical spectra, as detailed below.

Oshlack et al. deduce $M_{\rm BH}$ based on the H$\beta$ FWHM
and continuum luminosity. They find 16 RL AGN with
$M_{\rm BH}<10^8M_{\odot}$ (Table 1 there). The optical spectra
of four of these objects appear in Wilkes et al. (1983).
In PKS~1509+022
the spectrum is very red, the S/N is low, and H$\beta$ appears
to be significantly broader than estimated
(FWHM$> 10000$~km~s$^{-1}$, rather than 6550~km~s$^{-1}$),
implying $M_{\rm BH}>10^8M_{\odot}$ (rather than
$M_{\rm BH}=9.8\times 10^7M_{\odot}$). In PKS~1555$-$140 the
spectrum is also very red, there is no detectable
H$\beta$, and H$\alpha$ width is used instead. The
[S~II]$\lambda\lambda 6716, 6731$ doublet just redward of H$\alpha$
is remarkably strong and of comparable strength and width to
H$\alpha$, which suggests H$\alpha$ is mostly from the NLR.
In PKS~1725+044 the S/N is rather low, and H$\beta$ appears
to be significantly broader than estimated
(FWHM$\sim 8000$~km~s$^{-1}$, rather than 2400~km~s$^{-1}$,
implying $M_{\rm BH}\sim 8\times 10^8M_{\odot}$, rather than
$7\times 10^7M_{\odot}$).
In PKS~2143$-$156/R the FWHM estimate of 836~km~s$^{-1}$
is based on a low S/N Mg~II profile. This object was later
observed by Jackson \& Browne (1991a), who measured
a FWHM of 4700~km~s$^{-1}$ directly for H$\beta$
(Jackson \& Browne 1991b), implying
$M_{\rm BH}=1.5\times 10^9M_{\odot}$, rather than
$4.8\times 10^7M_{\odot}$. Thus in all four objects the
$M_{\rm BH}<10^8M_{\odot}$ determinations do not appear robust.

Of the remaining 12 $M_{\rm BH}<10^8M_{\odot}$ objects
in Oshlack et al., four have  $M_{\rm BH}<10^7M_{\odot}$.
These four objects also have the lowest luminosities,
$\lambda L_{\lambda}(5100)<1.2\times 10^{43}$~erg~s$^{-1}$, i.e.
$M_B>-19$. These very low luminosity AGN are further discussed
below. Of the remaining 8, only the spectrum
of PKS~2004$-$447 is published in Oshlack, Webster, \&
Whiting (2001). This object clearly has very narrow Balmer
lines from the BLR, although its SED is quite red. No published
optical spectra are available for the remaining 8 objects,
and the accuracy of the $M_{\rm BH}$ estimate could not be assessed.

\subsection{The Ho sample}

Ho (2002) studied the radio loudness vs. $M_{\rm BH}$ relation in
a heterogeneous sample of galaxies ranging over 12 orders of
magnitudes in $L_{H\beta}$. There are 16 RL AGN in this sample with
$M_{\rm BH}<10^8M_{\odot}$. Three of these are relatively luminous
having $M_B<-20.8$. The first (in Table 2 there) is 3C~120,
already discussed above; the second is PG~1211+143, which is
actually a RQQ (see corrected radio flux in Kellermann et al.
1994); and the third is PG 1704+608,
where the value of $M_{\rm BH}=3.7\times 10^7M_{\odot}$ is
taken from Kaspi et al., who used H$\beta$ FWHM=890 km~s$^{-1}$
which contrasts with the FWHM=6560 km~s$^{-1}$ obtained by Boroson \& Green
(1992). This large discrepancy results from an inaccurate
subtraction of the strong NLR H$\beta$ component, and the Boroson \& Green
value appears to be the correct one
(T. Boroson and S. Kaspi, private communications). Thus, the only luminous
($M_B<-20.8$~mag) RL AGN in the Ho sample at $M_{\rm BH}<10^8M_{\odot}$
is 3C~120, which is most likely beamed.

The remaining 13 RL AGN with $M_{\rm BH}<10^8M_{\odot}$
in the Ho sample all have $M_B>-17.4$~mag. There are indications that
low luminosity AGN have a different distribution of
$R$ parameters, compared to luminous AGN. In particular,
Ho \& Peng (2001) find that most of the very low
luminosity AGN ($M_B>-16$~mag, their Fig.4) are RL, and their $R$ value
distribution extends to higher values
($\sim 10^5$) than observed in high luminosity optically selected AGN
($R\sim 10^2-10^3$). Are all low luminosity AGN RL?
Does the radio loudness bimodality disappears at
low luminosities? The following section suggests it does not disappear.

\subsection{The luminosity dependence of $R$}

The Xu et al. study shows another remarkable
and apparently overlooked result. Their Fig.1 shows that the
bimodal distribution of AGN in the $L_{\rm 5 GHz}$ vs.
$L_{\rm [O III]}$ plane extends from the luminous AGN level
down by 4-5 orders of magnitude
in $L_{\rm [O III]}$. However, the dividing line which separates
the two populations follows $L_{\rm 5 GHz}\propto L_{\rm [O III]}^{\sim 0.5}$,
indicating that $R(\propto L_{\rm 5 GHz}/L_{\rm [O III]})$
along this line does not have a fixed value, but rather increases
towards low luminosity. Thus, the $R=10$ value commonly used to separate
RL from RQ AGN at high luminosities ($M_B\sim -26$,
e.g. Kellermann et al. 1989), is not the correct value to use at
low luminosity. The data of Xu et al. suggests that one should use
$R\propto L^{-0.5}$ to separate ``true" RL from ``true" RQ AGN as a
function of $L$. E.g., at $M_B\sim -16$ one should use $R\sim 10^3$
to separate the group of $R\sim 1-100$ RQ AGN from the group of
$R\sim 10^4-10^5$ RL AGN.

\section{Discussion}
\subsection{Common sources of error}
One major pitfall in existing $M_{\rm BH}$ estimates based on
the H$\beta$ FWHM
is proper subtraction of the NLR contribution to
H$\beta$. This problem tends to be significant in RLQ
compared to RQQ, since RLQ generally have a stronger NLR
component than RQQ at the same luminosity (e.g. Boroson \& Green 1992).
In some cases (mentioned above) the BLR
H$\beta$ component is either very weak or nonexistent, and only
the NLR H$\beta$ component was measured (but attributed to the BLR).
To properly subtract the NLR H$\beta$ component one
should typically assume
$f_{\rm H\beta}\simeq 0.1 f_{\rm [O~III]\lambda 5007}$,
the average ratio seen in NLR dominated AGN (e.g. Osterbrock 1989).

Another common source of error is very low S/N spectra, which can
result in highly inaccurate H$\beta$ FWHM values and $M_{\rm BH}$
estimates. Another thing to watch for is objects with a very red
continuum and an H$\alpha$/H$\beta$ flux ratio $>3$. In such
objects the H$\alpha$ line sometimes shows a dramatically different
profile from H$\beta$. In particular, H$\alpha$ may show
a strong broad component, which is either very weak or
undetectable in H$\beta$ (e.g. 1945+726, Stickel \& Kuhr 1993a).

\subsection{The apparently robust cases}
The apparently robust cases of RL AGN with
$M_{\rm BH}<10^8M_{\odot}$ are found in two populations
of objects. First, compact flat spectrum sources,
such as PKS~2004-447 (Oshlack et al. 2001, 2002) and
3C~120. As discussed by Jarvis \& McLure (2002), their
radio flux is most likely enhanced by beaming, and their
observed H$\beta$ line width may be biased to low values
if the BLR is in a face on disk. However, it is difficult to make
an accurate quantitative correction to both effects,
especially if large corrections are expected.
Independent estimates of $M_{\rm BH}$ in compact flat spectrum
sources can be obtained from measurements of the host galaxy
luminosity and the $M_{\rm BH}$ - bulge luminosity relation.
Preliminary results based on this method  suggest luminous host
galaxies and thus large $M_{\rm BH}$
in the Oshlack et al. objects (Jarvis \& McLure 2003).

Apparently, the only robust case of low $M_{\rm BH}$ RL AGN consists of
low luminosity AGN (Ho 2002). However, here we may be missled
by using $R=10$ to separate RL from RQ AGN, which applies for
luminous AGN. The large compilation of Xu et al. suggests
that $R\sim 1000$ may be a more relevant number when going down
from $L\sim 10^{46}$~erg~s$^{-1}$
AGN to $L\sim 10^{42}$~erg~s$^{-1}$ AGN. With this revised, luminosity
dependent radio loudness $R$ threshold, practically all of the
$M_{\rm BH}<10^8M_{\odot}$ AGN in the sample of Ho belong to the
``true" RQ AGN population.

\subsection{The $R$-$M_{\rm BH}$ relation}
The various literature results pointed out in this comment suggest
that {\em the radio loudness vs. $M_{\rm BH}$ relation extends to
low luminosity AGN} . Specifically, that all true  RL AGN have
$M_{\rm BH}>10^8M_{\odot}$, and all AGN with
$M_{\rm BH}<10^8M_{\odot}$ are true RQ AGN. An additional
interesting hint supporting this suggestion can be seen in Figure 5 of
Xu et al., which shows the distribution of host galaxies of the AGN
in the $L_{\rm 5 GHz}$ vs.
$L_{\rm [O III]}$ plane. Spiral hosts populate only the lower ``true"
RQ AGN branch, and only Elliptical and S0 hosts appear in the upper RL
AGN branch.

\subsection{The $R$ bimodality}
Finally, the lack of radio loudness bimodality in the FIRST survey
appears to be consistent with a similar lack of bimodality in known
samples when one uses the core, rather than total, radio emission.
Since the extended radio emission probably
provides a better estimate of the intrinsic radio power, the radio
emission in AGN is likely to be intrinsically bimodal.

\section{Appendix}
For the sake of convenience we reproduce here some of the relevant
figures mentioned in this paper. Higher quality versions are
available in the original publications. \\[-2cm]
\begin{figure}[h]
\plottwo{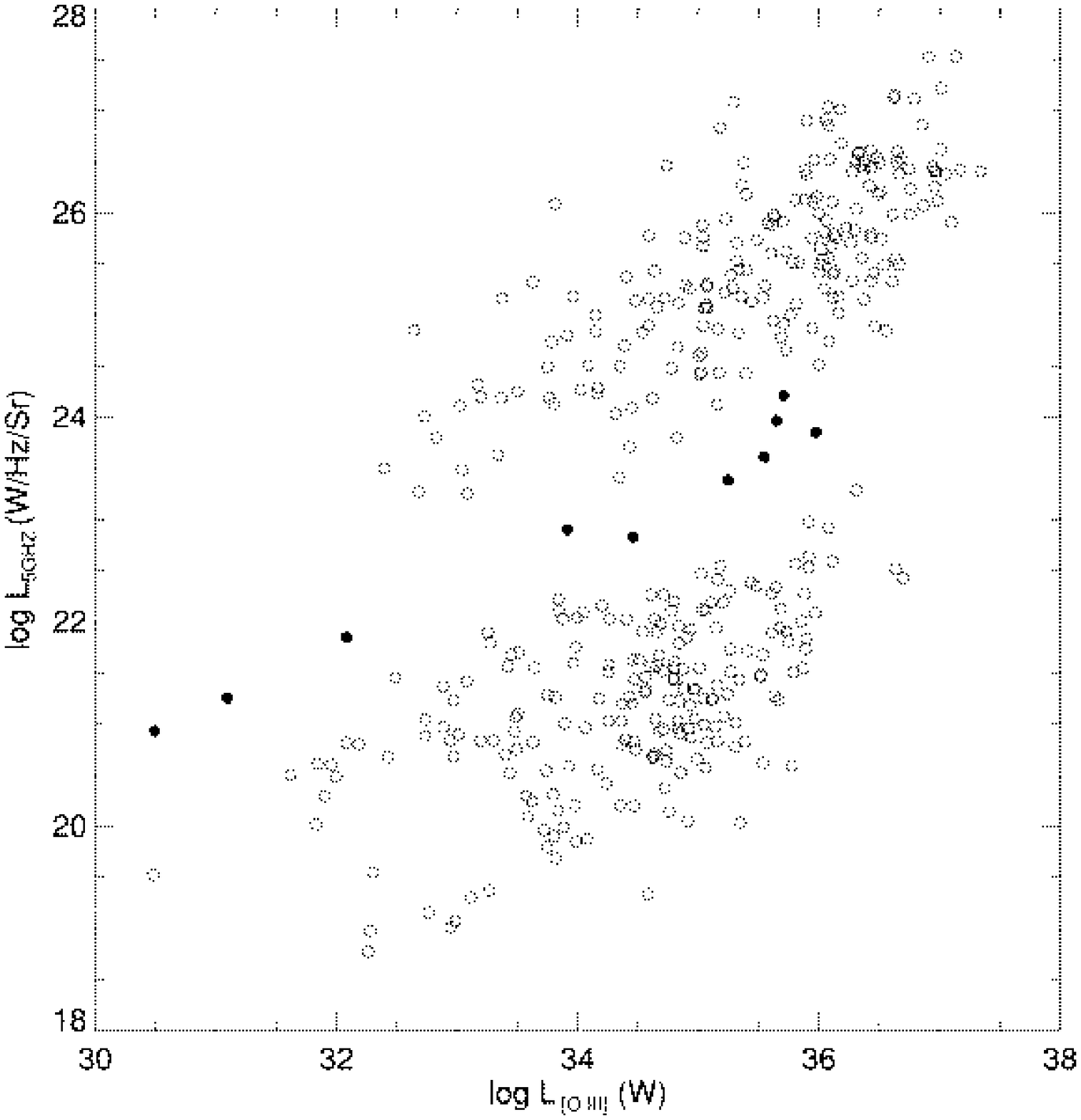}{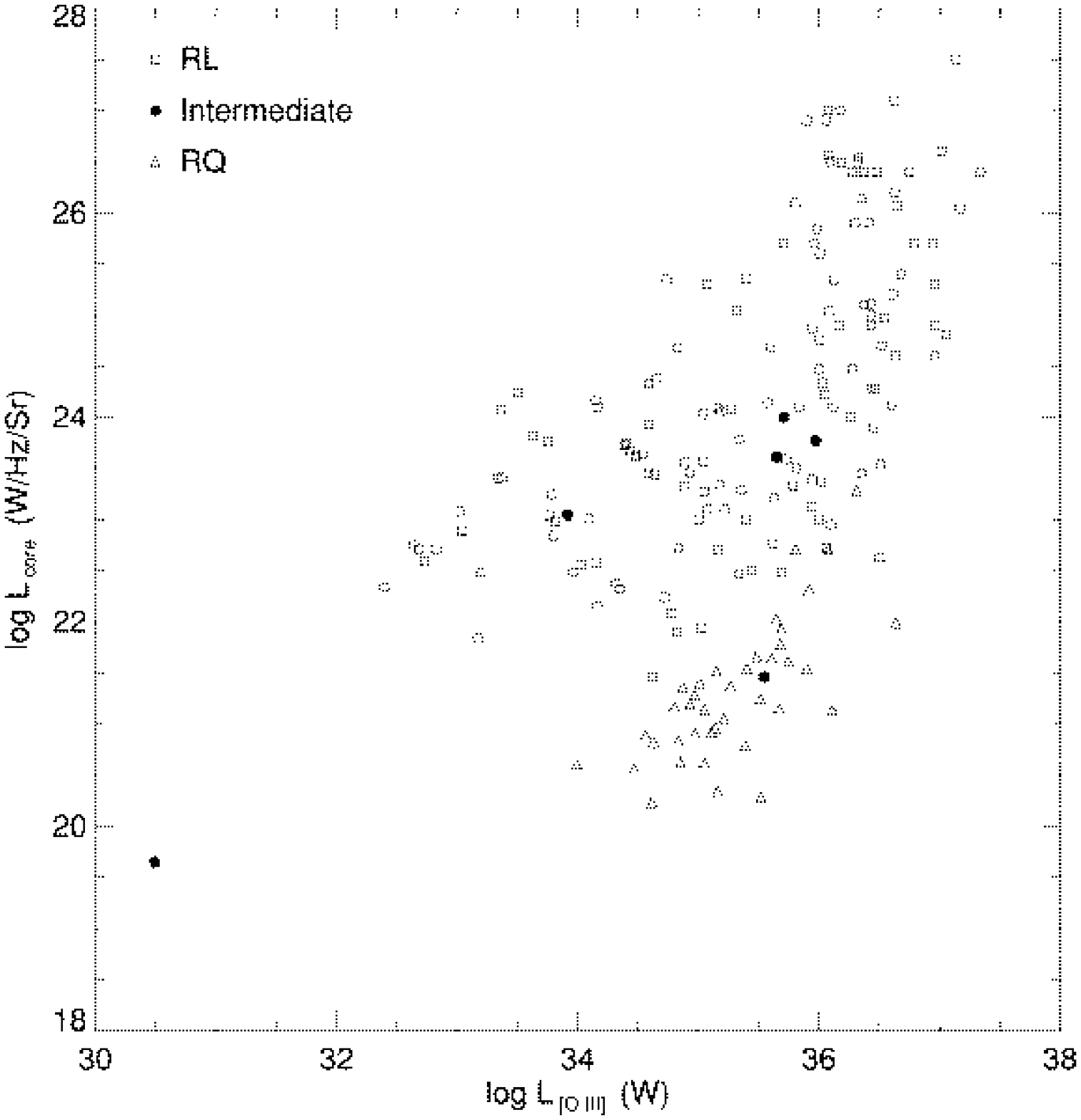}
\caption{Left panel: Figure 1 from Xu et al. (1999). Note that the radio loudness
bimodality extends to low $L_{[O III]}$, and that the seprating
$L_{5GHz}\propto L_{[O III]}^{0.5}$.
 Right panel: Figure 3 from Xu et al. Note that the bimodal distribution
 disappears when only the core 5GHz luminosity is used.}
\plottwo{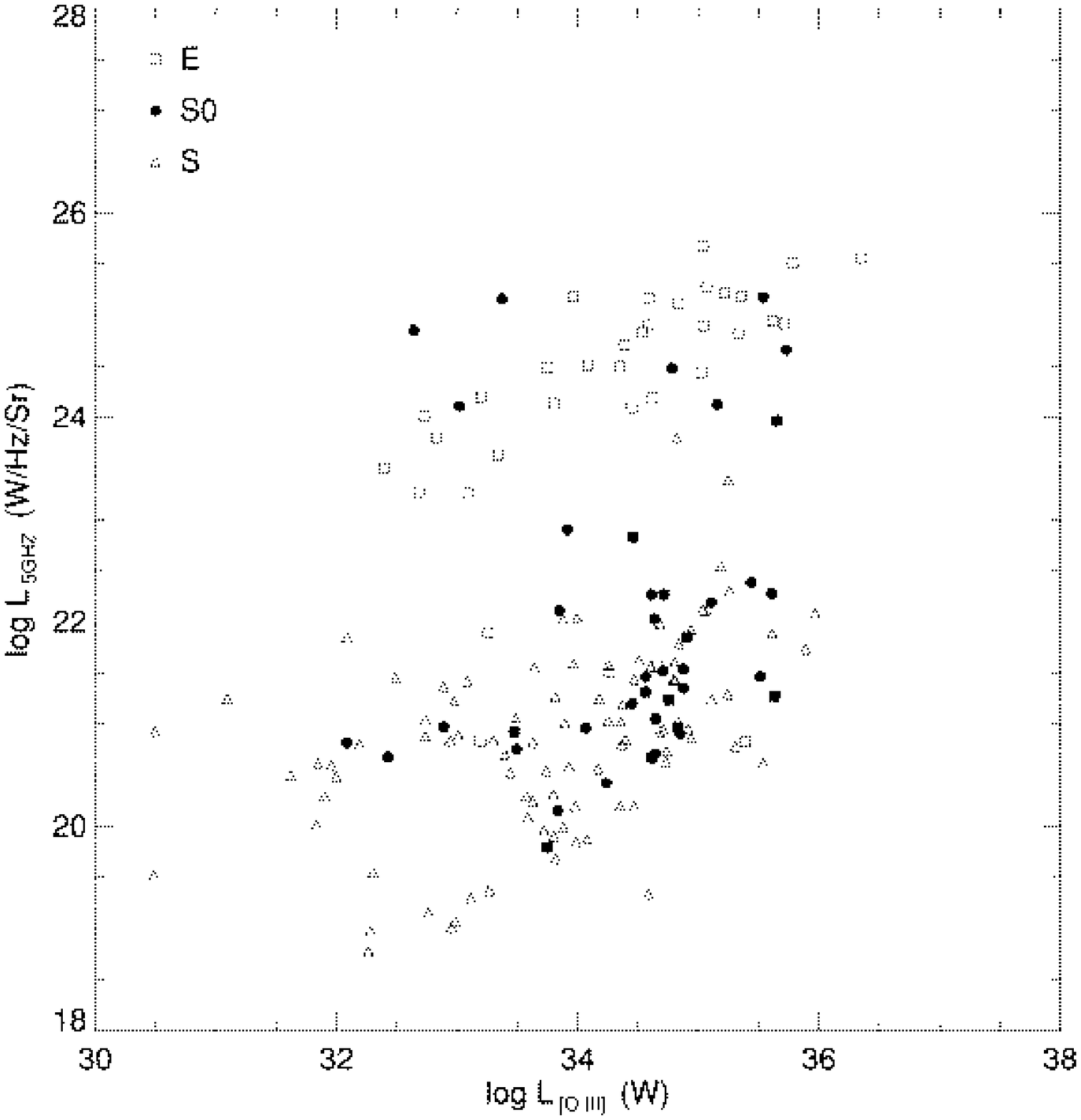}{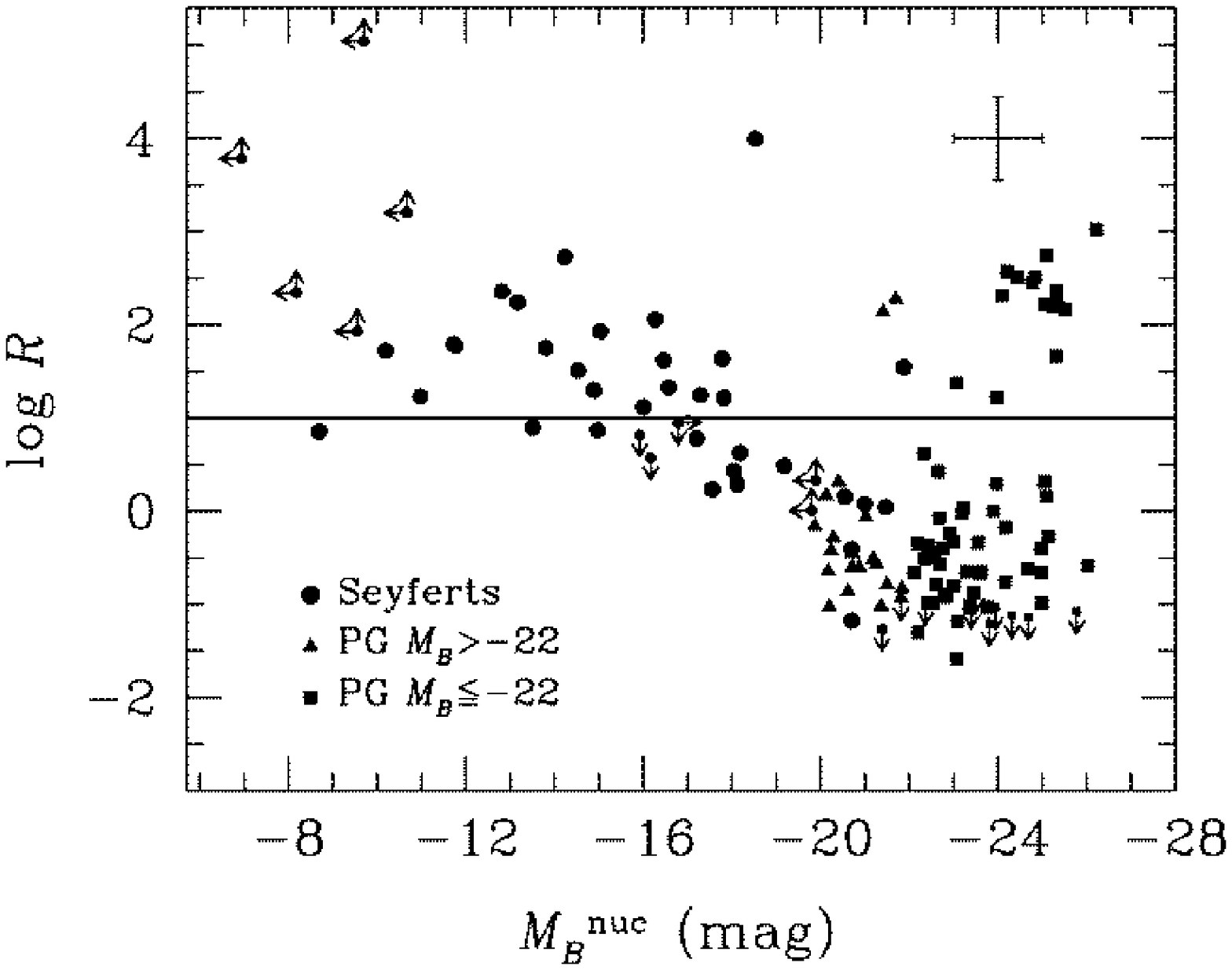}
\caption{Left panel: Figure 5 from Xu et al. Note that the host morphology
radio loudness relation is maintained down to low $L_{[O III]}$, which
hints that the radio loudness - $M_{\rm BH}$ relation is also
maintained.
Right panel: Figure 4 from Ho \& Peng (2001). Note that most AGN at
$M_B>-16$ appear to be RL.}
\end{figure}
\begin{figure}[h]
\plottwo{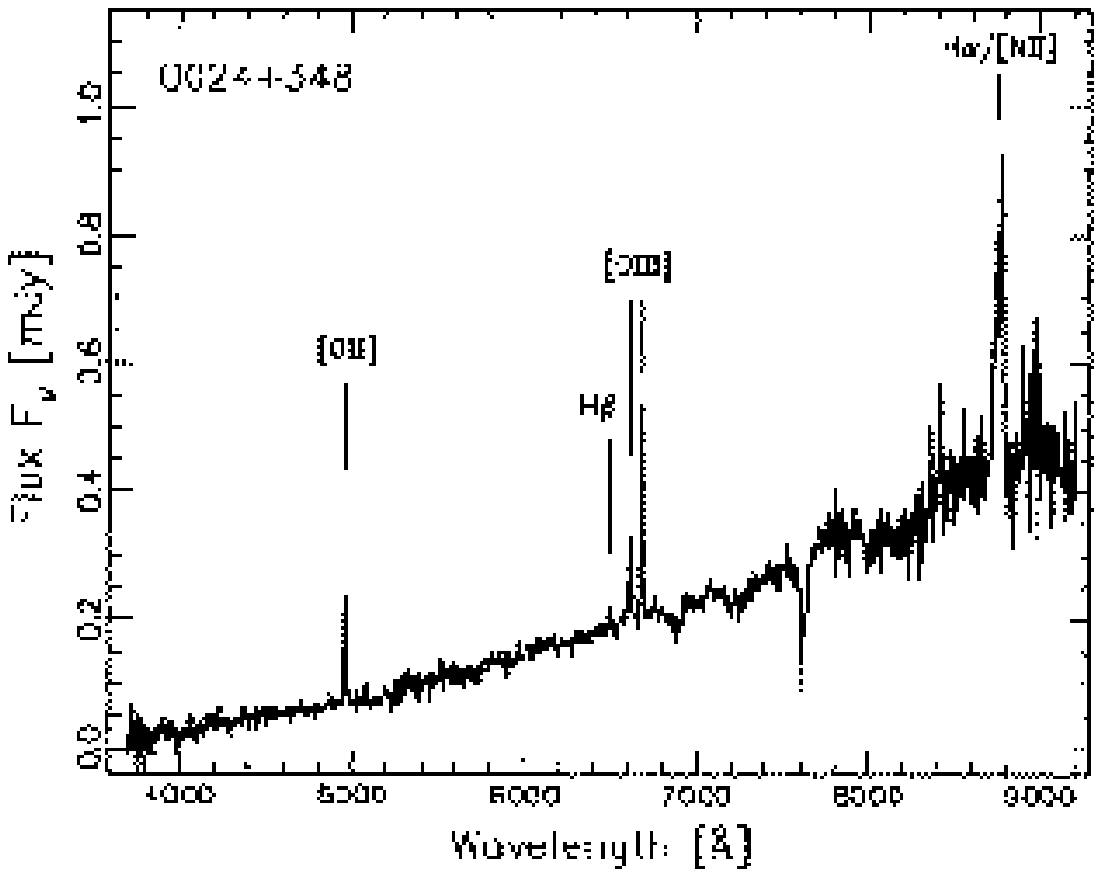}{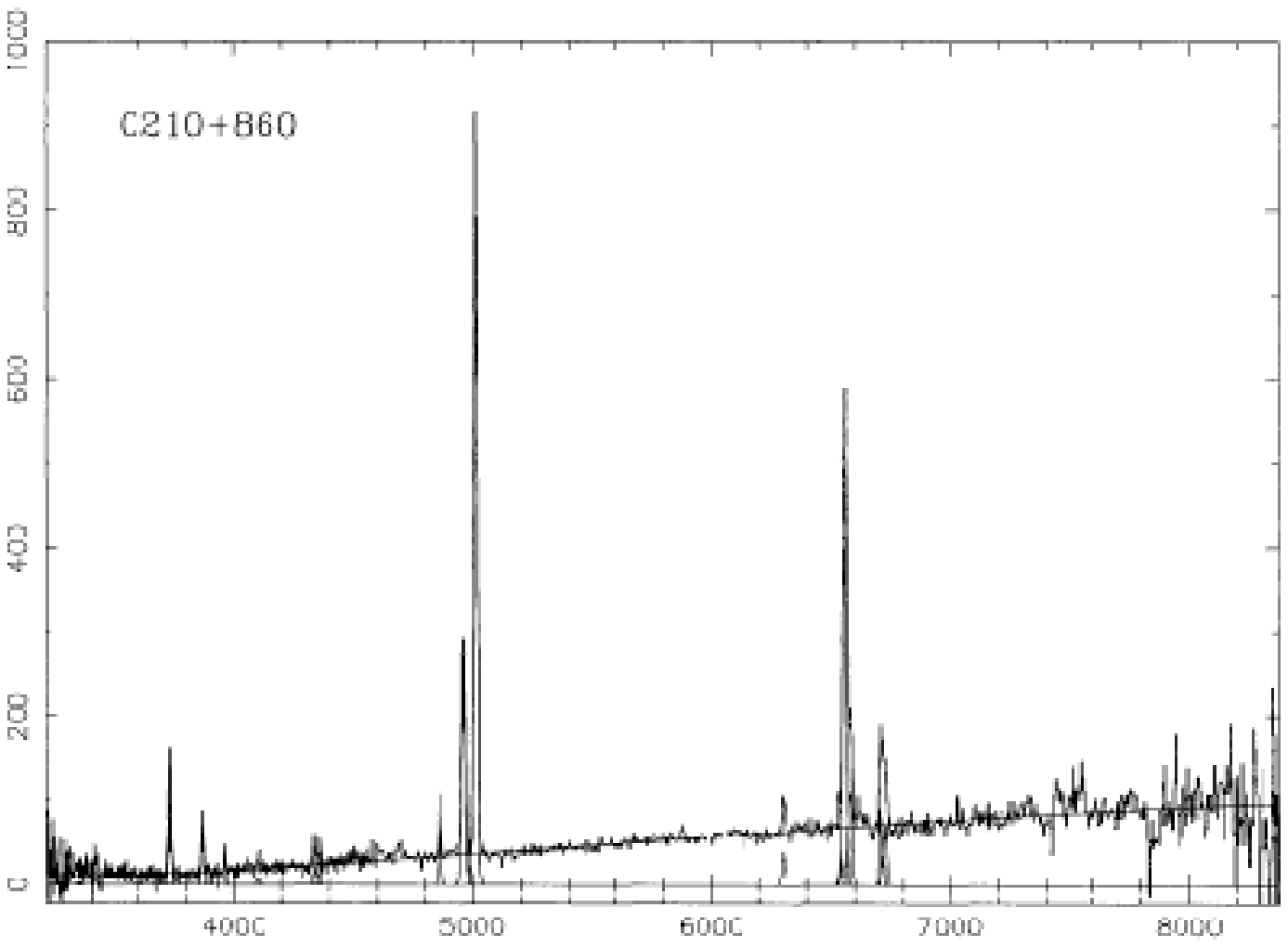}
\caption{Left panel: 0024+348 (Stickel \& Kuhr 1993b).
Right panel: 0210+860 (Lawrence et al. 1996).}
\plottwo{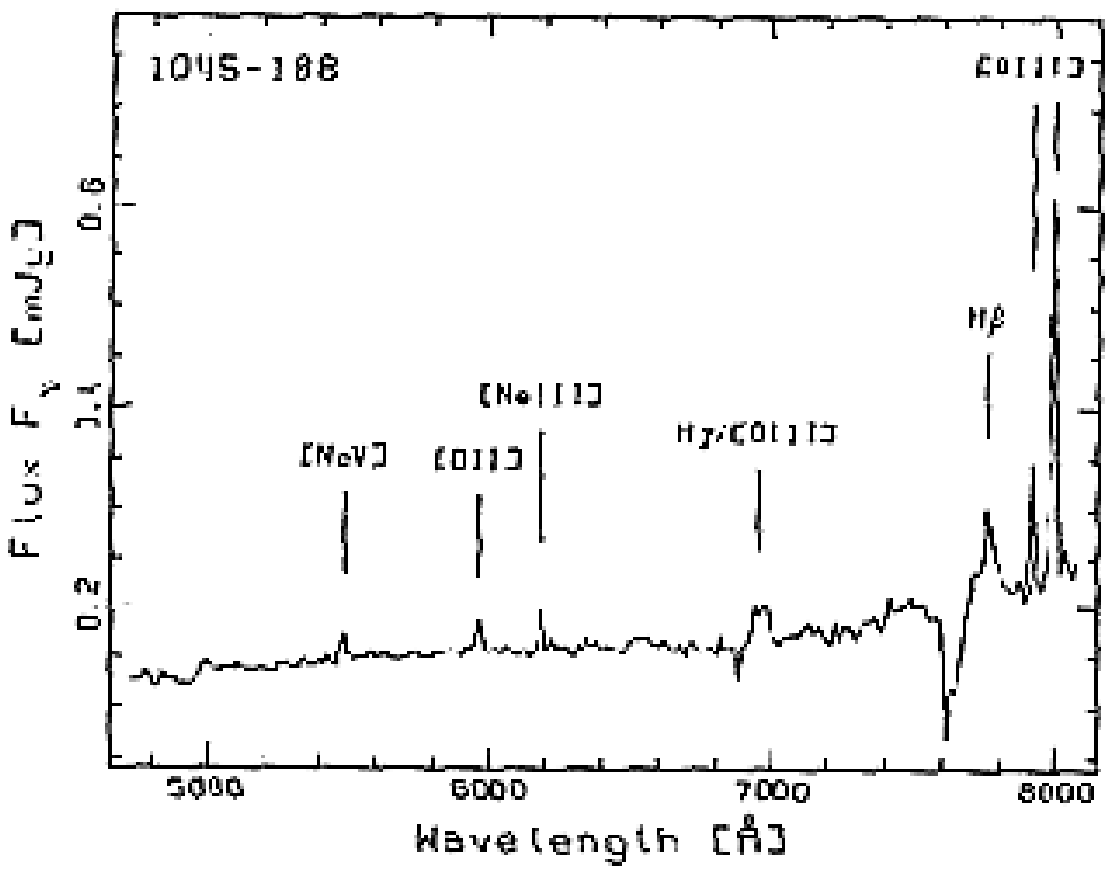}{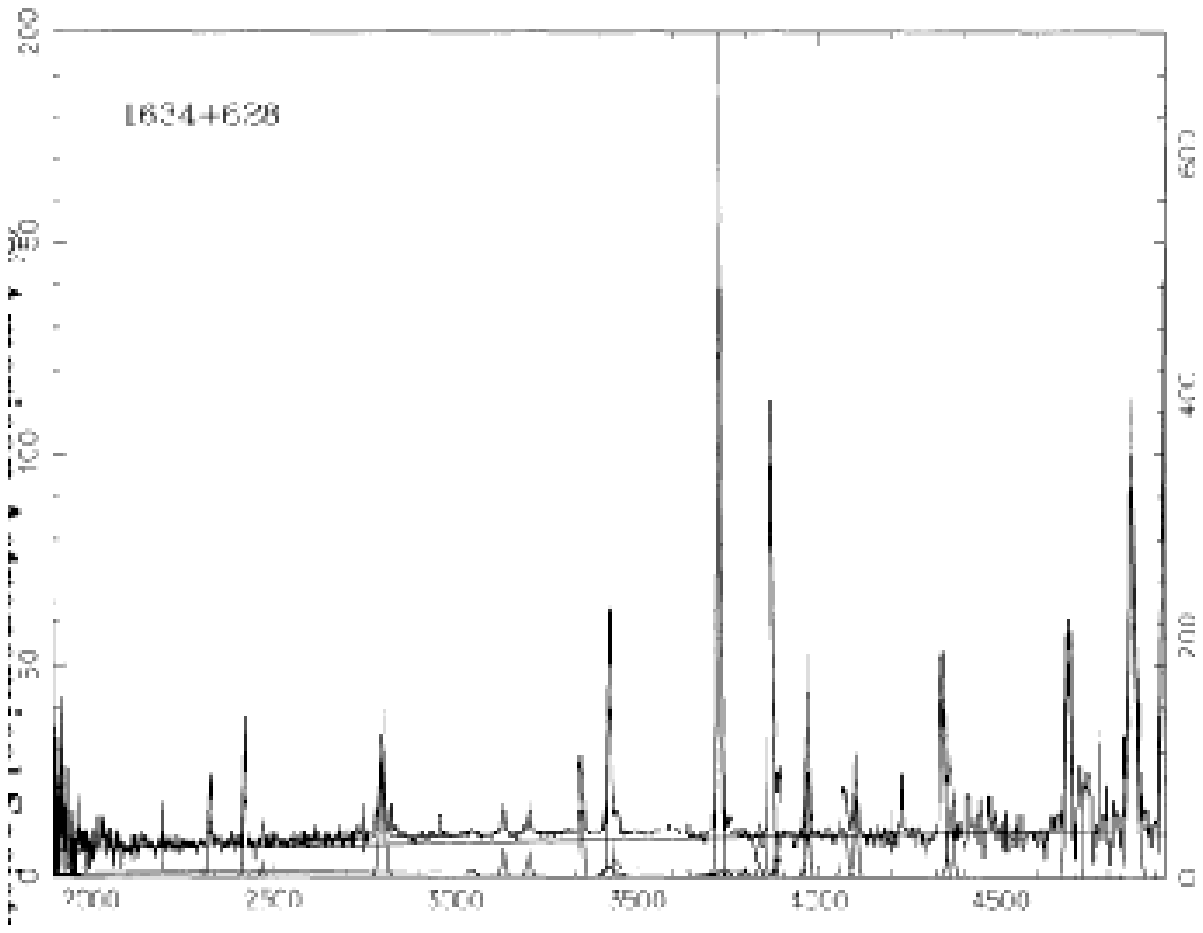}
\caption{Left panel: 1045-188 (Lawrence et al. 1996).
Right panel: 1634+628 (Lawrence et al. 1996).}
\plottwo{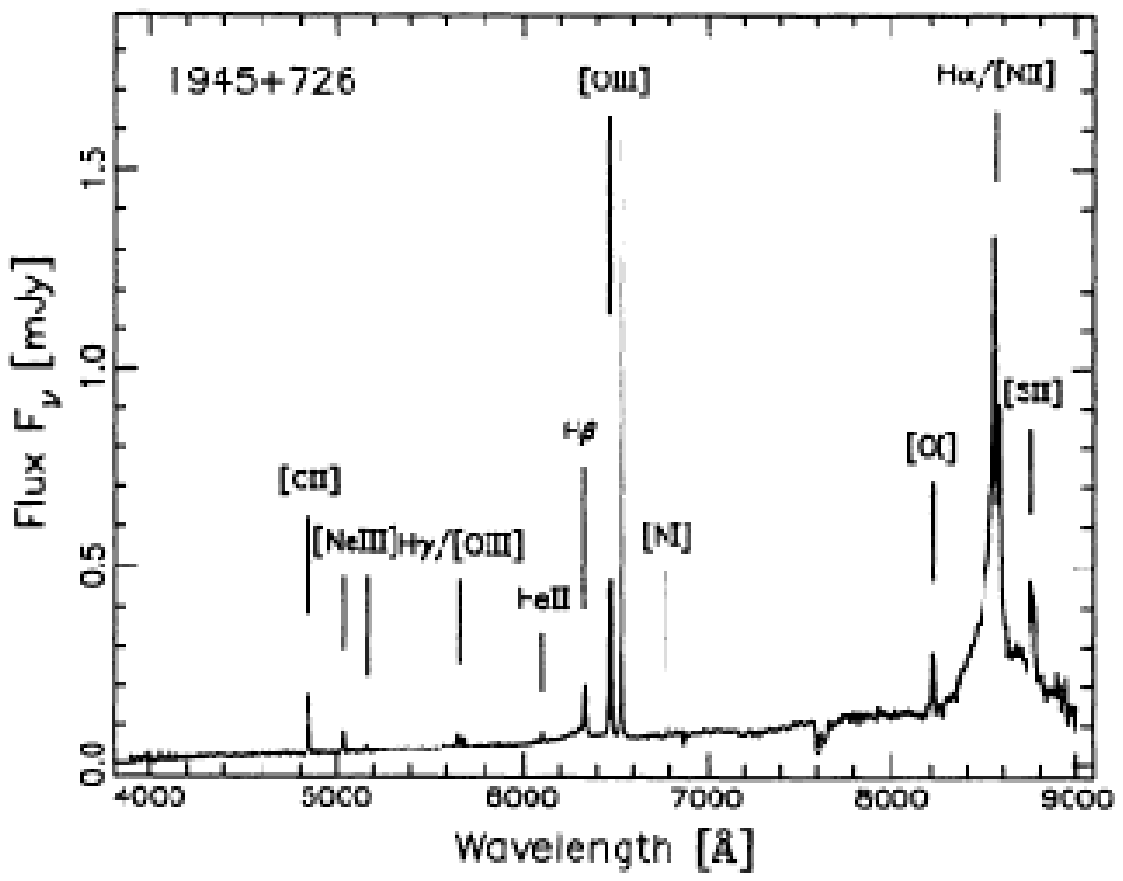}{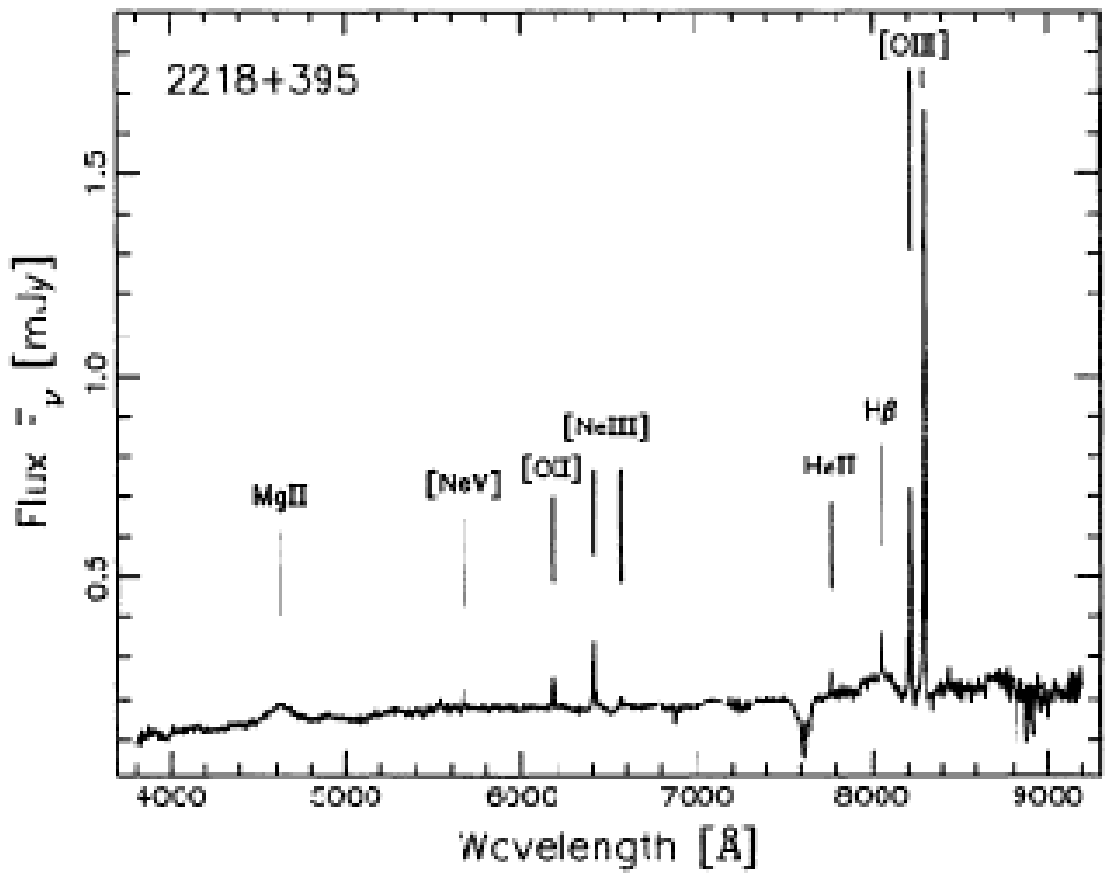}
\caption{Left panel: 1945+725 (Stickel \& Kuhr 1993a).
 Right panel: 2218+395 (Stickel \& Kuhr 1993b)}
\end{figure}
\begin{figure}
\plotfiddle{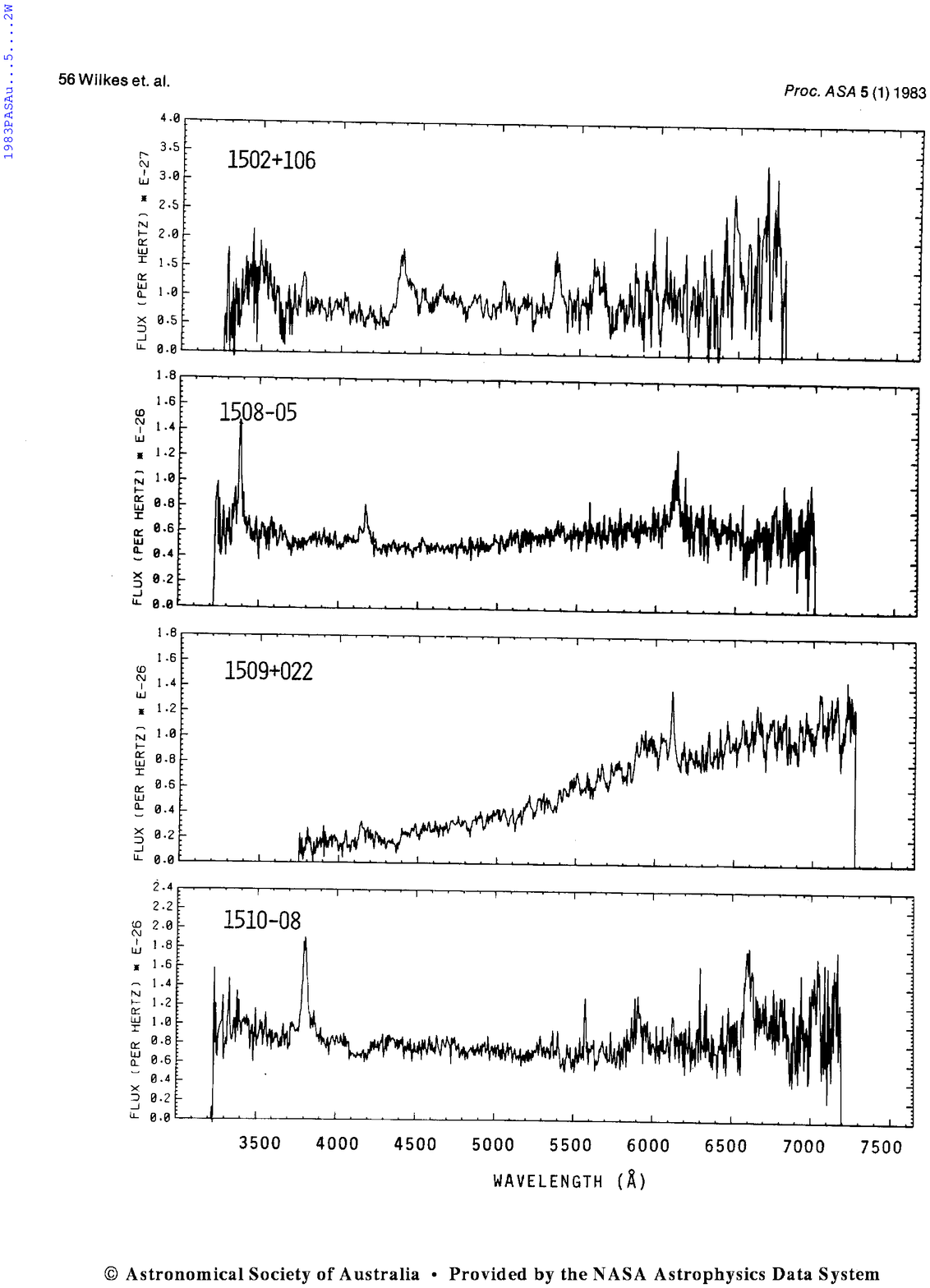}{21.5cm}{0.6}{80}{80}{0}{0}
\caption{The spectrum of PKS 1509+022 on the third panel
(Wilkes et al. 1983). Note the wavelength scale, which
also applies for the following plots from Wilkes et al.}
\end{figure}
\begin{figure}[h]
\plottwo{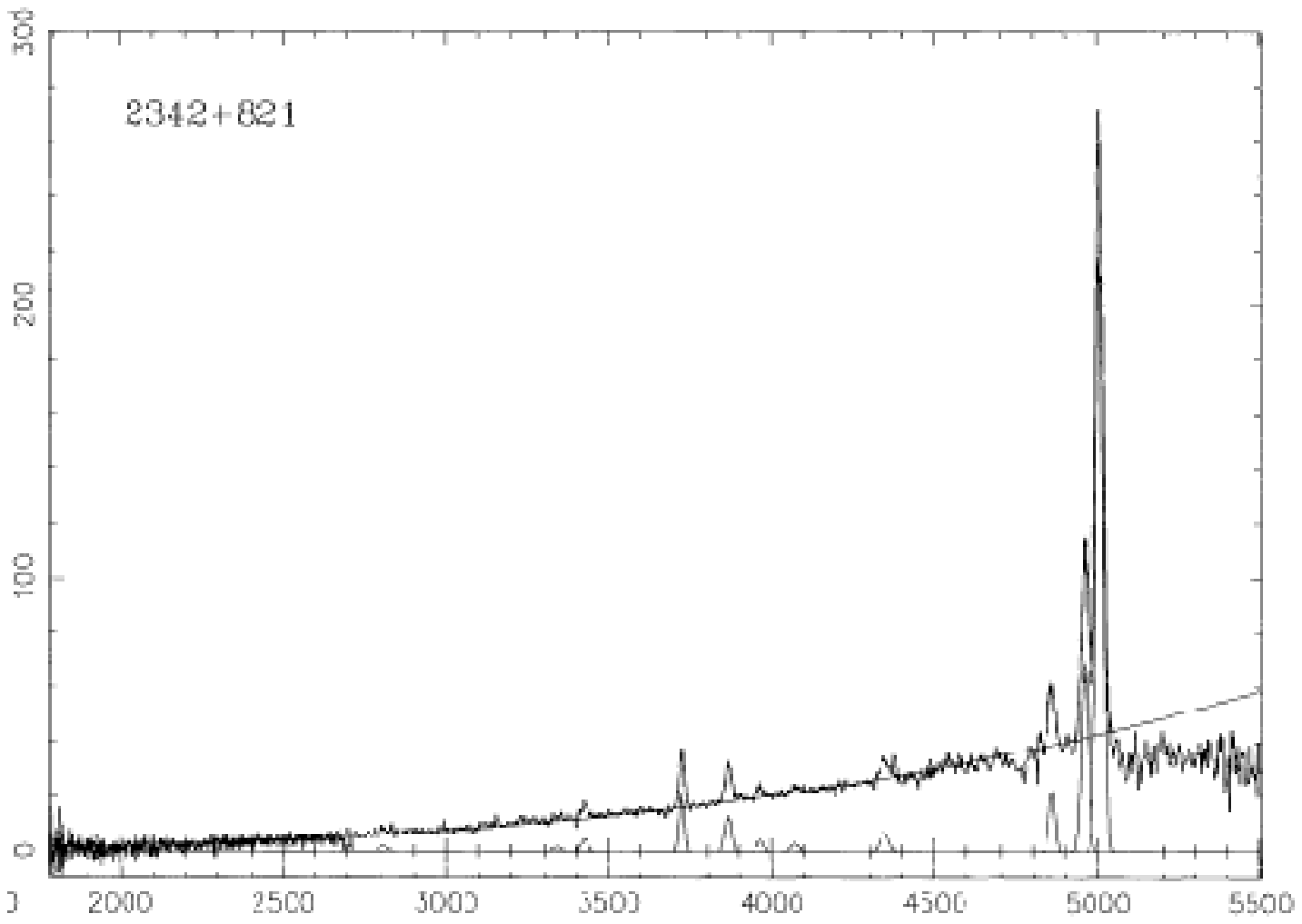}{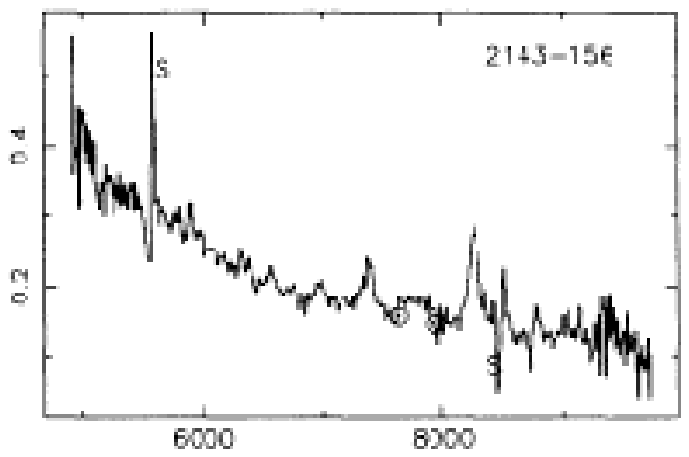}
\caption{Left panel: 2342+821 (Lawrence et al. 1996).
Right panel: PKS 2143+156 (Jackson \& Browne 1991).}
\plotone{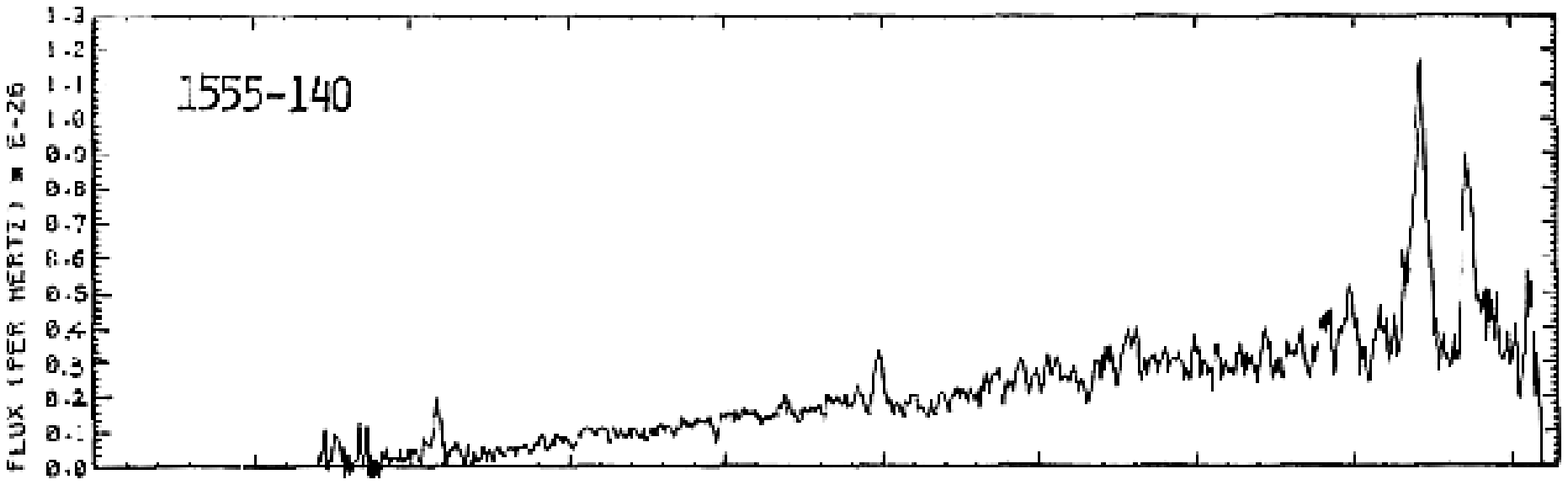}
\caption{PKS 1555$-$140 (Wilkes et al. 1983)}.
\plotone{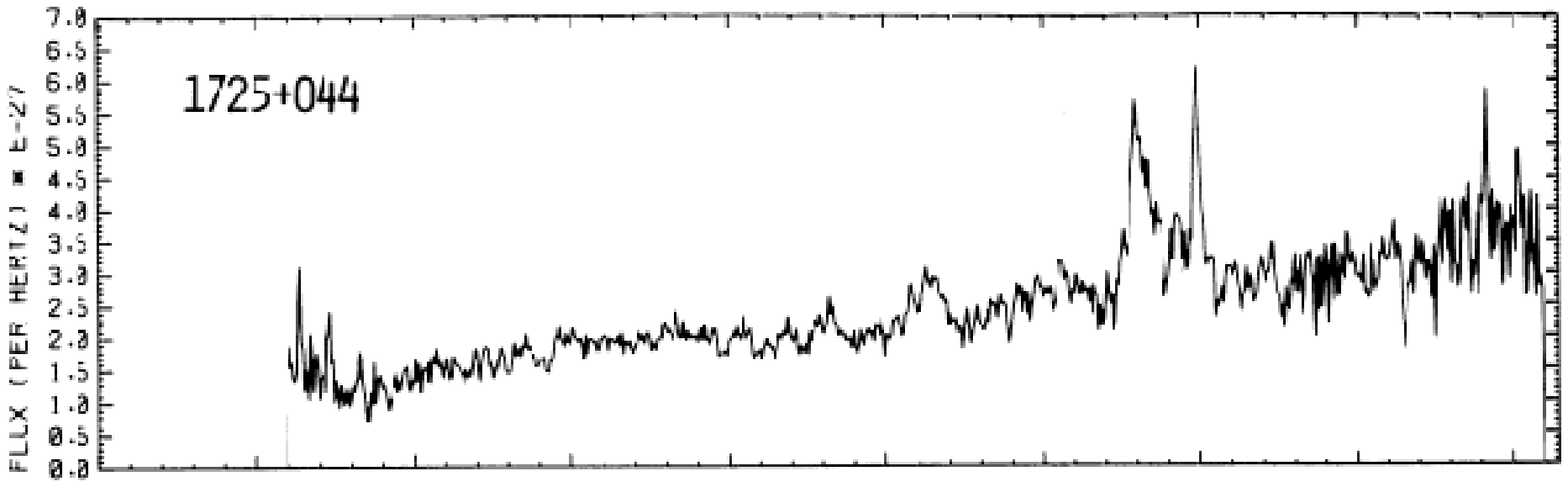}
\caption{PKS 1725+044 (Wilkes et al. 1983)}.
\plotone{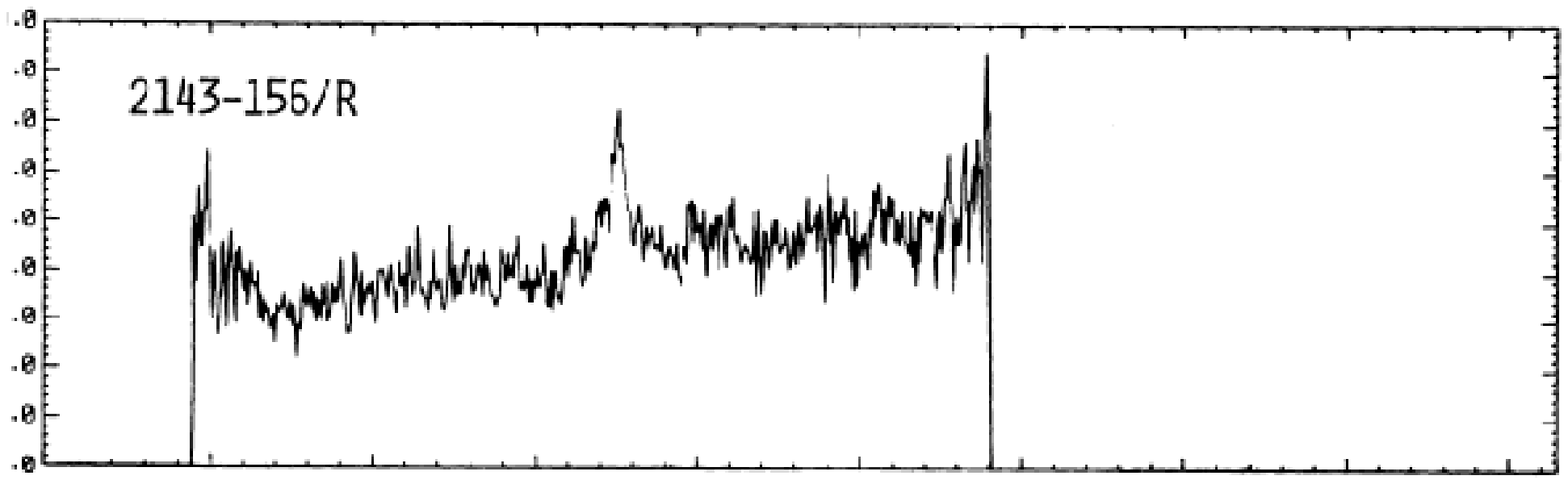}
\caption{PKS 2143-156 (Wilkes et al. 1983)}.
\end{figure}

\end{document}